# Size-independent susceptibility to transport in aeolian saltation

Raleigh L. Martin[a*], Jasper F. Kok[a]

[a]Department of Atmospheric and Oceanic Sciences, University of California, Los Angeles, CA 90095, USA.

[*]**Corresponding Author**
Raleigh L. Martin
Department of Atmospheric and Oceanic Sciences
University of California, Los Angeles
520 Portola Plaza
Los Angeles, CA 90095
e-mail: raleighmartin@gmail.com

**Running title:** Equal susceptibility in aeolian saltation

**Key Points**
- A single threshold governs the susceptibility of most surface particle size classes to saltation
- Equal susceptibility can be explained by the splash mechanism that entrains most particle size classes
- Non-interacting single particle treatments are problematic; using a reference grain size could improve sand flux and dust emission models

## Abstract
Natural wind-eroded soils contain a mixture of particle sizes. However, models for aeolian saltation are typically derived for sediment bed surfaces containing only a single particle size. To nonetheless treat natural mixed beds, models for saltation and associated dust aerosol emission have typically simplified aeolian transport either as a series of non-interacting single particle size beds or as a bed containing only the median or mean particle size. Here, we test these common assumptions underpinning aeolian transport models using measurements of size-resolved saltation fluxes at three natural field sites. We find that a wide range of sand size classes experience "equal susceptibility" to saltation at a single common threshold wind shear stress, contrary to the "selective susceptibility" expected for treatment of a mixed bed as multiple single particle size beds. Our observation of equal susceptibility refutes the common simplification of saltation as a series of non-interacting single particle sizes. Sand transport and dust emission models that use this incorrect assumption can be both simplified and improved by instead using a single particle size representative of the mixed bed.

# 1. Introduction

The particle size distribution (PSD) of a soil surface affects the ability of the wind to erode particles and to transport them in aeolian saltation, i.e., the ballistic hopping of wind-blown sand grains (e.g., Chepil, 1945a; Iversen et al., 1976). Determining soil PSD effects on saltation dynamics is important for understanding a range of processes governed by aeolian saltation, including the migration of sand dunes (e.g., Gao et al., 2016), the recording of wind conditions in aeolian stratigraphy (e.g., Lindhorst and Betzler, 2016), the erosion of exposed agricultural soils (e.g., Chepil, 1945a), and the emission of airborne dust (e.g., Alfaro and Gomes, 2001; Shao, 2001) and consequent effects on climate (e.g., Gillette and Passi, 1988; Kok et al., 2017).

Although the dynamics of saltation are relatively well understood for idealized single particle size (i.e., homogeneous, well-sorted, or monodisperse) soil beds (e.g., Bagnold, 1941; Durán et al., 2011; Kok et al., 2012), there is currently no clear understanding of how the presence of other particle sizes in natural mixed (i.e., heterogeneous, poorly-sorted, or polydisperse) soils affects both the threshold wind speed to initiate particle motion and the subsequent flux of a given particle size class (Williams, 1964; Gillette and Walker, 1977). In the absence of comprehensive field measurements to inform saltation dynamics over mixed soils, some models of sand transport and dust emission have assumed that particles of different sizes saltate independently of each other (Marticorena and Bergametti, 1995; Shao et al., 1996; Alfaro and Gomes, 2001; Zender et al., 2003), and thus that the saltation flux of a given particle size class is unaffected by the presence of other size classes. Other models assume that a single grain size (i.e., median or mean soil particle size) can be used to represent all particle size classes on the bed (e.g., Elbelrhiti et al., 2005; Andreotti et al., 2010). These respective 'independent' (Shao, 2008, p. 174) and 'representative' (e.g., Claudin and Andreotti, 2006, p. 40) saltation assumptions have not been tested, which is potentially problematic, as mixed soils dominate the coastal (e.g., Greeley et al., 1996), desert (e.g., Gillette et al., 1980), agricultural (e.g., Gillette, 1988), and planetary (e.g., Claudin and Andreotti, 2006) settings where aeolian processes take place.

The potential size-selectivity of aeolian saltation over a mixed soil bed can be broken down into two primary components, which we term "susceptibility" and "mobility" (Fig. 1). "Susceptibility" describes the threshold shear stress required to entrain particles of a certain size class from the bed into saltation. "Equal susceptibility" therefore refers to the case in which all grain size classes are subject to the same threshold stress, whereas "selective susceptibility" refers to a variation in threshold with grain size. "Mobility" describes the possible variation in the proportionality between saltation flux and the above-threshold shear stress for a specific size class; thus, "equal mobility" refers to all size classes governed by the same stress-flux proportionality and "selective mobility" refers to a size-selective saltation transport relationship. (To dispel possible confusion, we note here that analogous studies of fluvial bedload sediment transport have used the term "equal mobility" somewhat inconsistently (e.g., Parker and Toro-Escobar, 2002), possibly referring to either equal susceptibility or equal mobility as we define them here.)

We can formalize these definitions of equal and selective susceptibility and mobility by considering the flux law relating saltation flux to wind shear stress. Considering first the bulk (non size-selective) relationship between saltation mass flux $Q$ [gm$^{-1}$s$^{-1}$] and wind shear stress $\tau$

[Pa] ($= \rho_f u_*^2$, where $\rho_f$ is air density and $u_*$ is wind shear velocity [ms$^{-1}$]), it is shown by theory (e.g., Ungar and Haff, 1987), wind tunnel experiments (e.g., Ho et al., 2011), and recent field measurements (Martin and Kok, 2017) that the saltation flux law takes the form:

$$Q = C(\tau - \tau_{th}), \tag{1}$$

where the flux coefficient $C$ [s] and the threshold stress $\tau_{th}$ [Pa] both depend on soil and air properties (e.g., Kok et al., 2012; Martin and Kok, 2017).

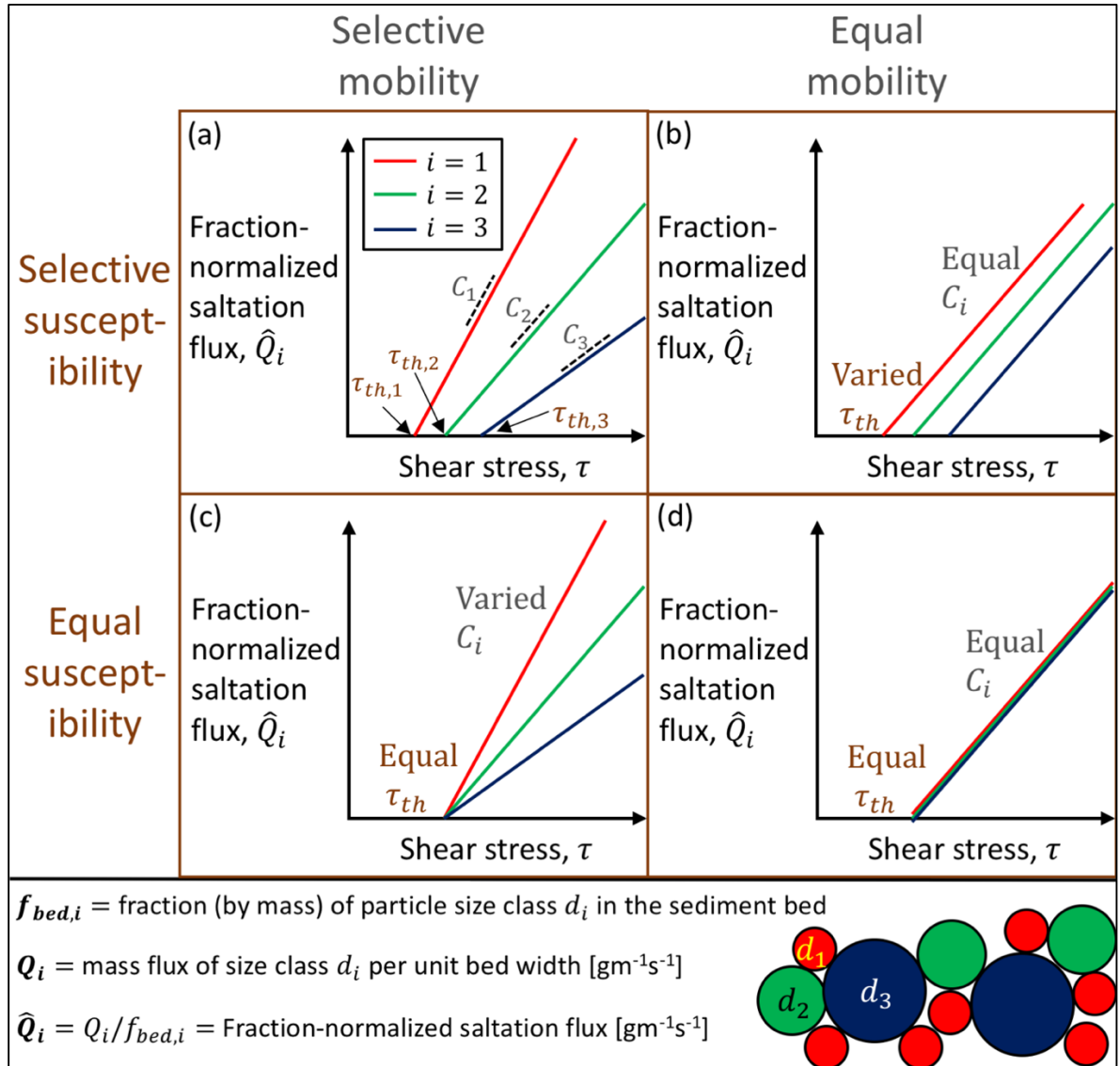

**Figure 1.** Conceptual diagram illustrating the differences between selective and equal susceptibility and mobility for a sediment bed composed of three distinctive particle size classes, $d_1 - d_3$. In the case of "selective susceptibility", $\tau_{th,i}$ varies with each $d_i$ (panels a and b). In contrast, for "equal susceptibility" (panels c and d), all of the $d_i$ are governed by a single $\tau_{th}$. In addition to differences in susceptibility, the mobility of particles can also be size selective. In the case of "selective mobility" (panels a and c), the value of $C_i$ varies with $d_i$. In contrast, for "equal mobility" (panels b and d), all of the $d_i$ are governed by a single $C$.

Now, for a sand bed containing a heterogeneous mixture of grain sizes, we consider distinctive particle size classes $d_i$, each constituting a fraction $f_{bed,i}$ of sediments exposed on the surface of the sediment bed. We then define $Q_i$ as the saltation mass flux for this specific particle size class, and we define the fraction-normalized size-selective saltation flux as $\hat{Q}_i = Q_i/f_{bed,i}$. We can then define a size-selective saltation flux law as:

$$\hat{Q}_i = C_i(\tau - \tau_{th,i}), \qquad (2)$$

where $C_i$ and $\tau_{th,i}$ describe the size-selective saltation flux coefficient and threshold shear stress, respectively. As illustrated in Fig. 1, the distinction between equal and size-selective mobility and susceptibility depends on whether $C_i$ and $\tau_{th,i}$ vary with particle size class $d_i$ in Eq. 2. In particular, selective and equal susceptibility are distinguished by whether or not $\tau_{th,i}$ varies with $d_i$; selective and equal mobility are distinguished by the presence or absence of variation in $C_i$.

Our limited existing understanding of size-selectivity in the susceptibility and mobility of aeolian saltation derives from measurements of the PSD of airborne saltators. Wind tunnel (Williams, 1964; Gillette and Walker, 1977), field (Kok et al., 2012, based on field data of Namikas, 2003), and numerical (Kok and Renno, 2009) studies indicate that this airborne PSD is mostly insensitive to wind shear stress. This suggests a common threshold shear stress for the initiation of aeolian saltation for all size classes – that is, equal susceptibility – which conflicts with the 'independent saltation' assumption made in most sand transport and dust emission models (e.g., Marticorena and Bergametti, 1995; Shao, 2008). Field (Gillette and Walker, 1977) and numerical (Kok and Renno, 2009) studies also indicate a slight fining of the airborne PSD with respect to the surface PSD, while wind tunnel studies show either negligible (Xing, 2007; Li et al., 2008) or substantial (Williams, 1964) fining of the airborne PSD. Therefore, aeolian saltation probably displays some degree of selective mobility, but the extent of this size-selectivity is unclear.

Here we examine the size-selectivity of both susceptibility and mobility in aeolian saltation by analyzing size-resolved saltation measurements at multiple field sites. After first describing methods for discerning size-selective saltation properties from measured PSDs and saltation fluxes, we present results indicating equal susceptibility in aeolian transport. We also briefly examine size-selective mobility, though our results are limited by a lack of near-surface saltation measurements. Finally, we consider implications for treating saltation processes over mixed soils in geomorphic and dust emission models, and we address limitations in the common assumption of independent saltation of different grain sizes for mixed grain-size sediment beds.

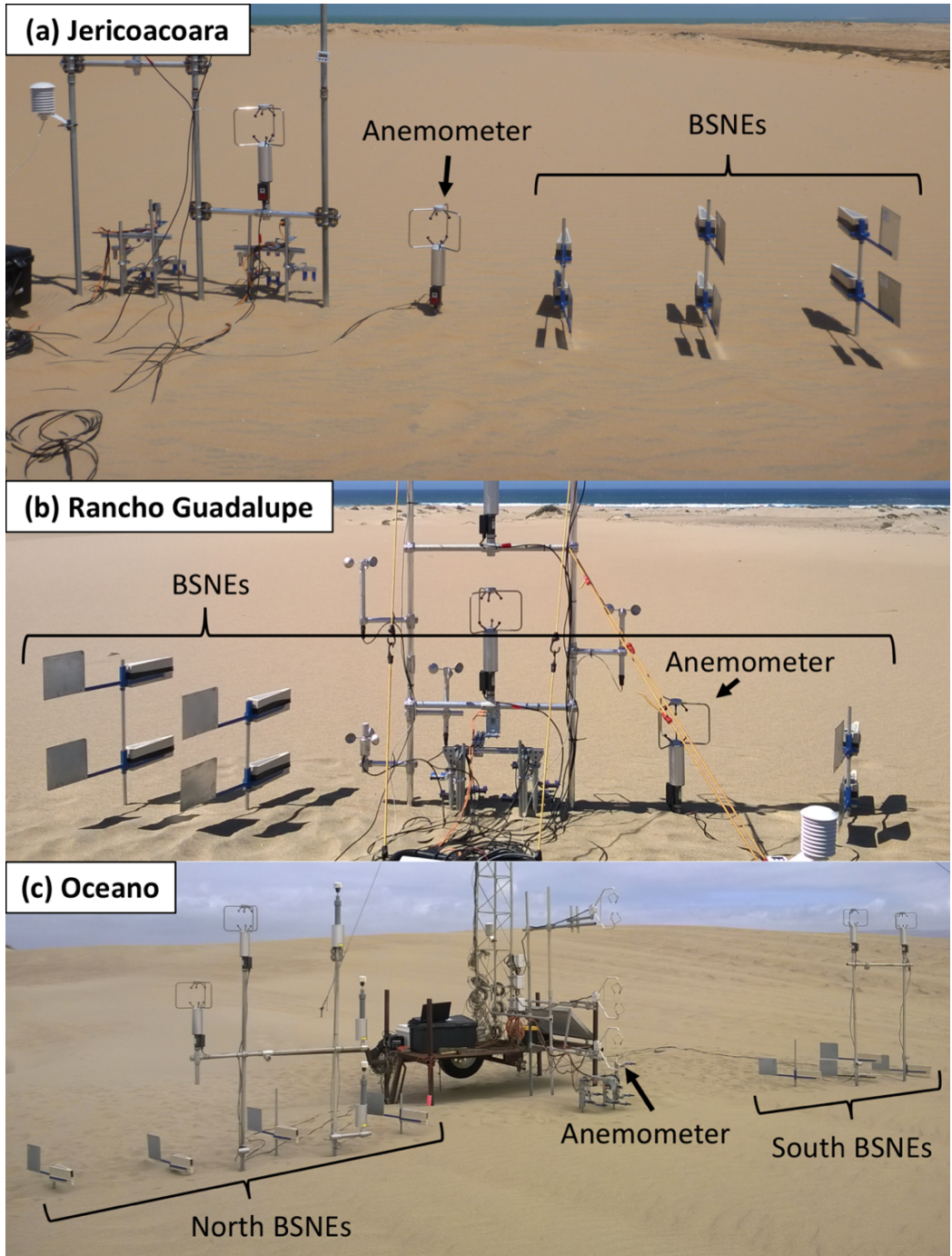

**Figure 2.** Photos illustrating configurations of BSNE traps and anemometers at three locations for size-selective transport analyses: (a) Jericoacoara (looking upwind), (b) Rancho Guadalupe (looking upwind), and (c) Oceano (looking downwind). Though multiple instruments were deployed at each site, our analyses depend only on wind measurements from a single anemometer mounted at $z \approx 0.5$m (indicated in each panel) and BSNE saltation traps. At Jericoacoara, maximum spanwise separation among BSNEs was $\approx$1.5m. At Rancho Guadalupe, maximum BSNE separation was $\approx$3.6m. At Oceano, BSNEs were separated into a "North" (N) and "South" (S) cluster. The distance between the cluster centers was $\approx$8m; maximum spacing within the N and S clusters was $\approx$2.9m and $\approx$2.1m, respectively.

**Table 1.** The median ($d_{50,bed}$), mean ($\bar{d}_{bed}$), modal ($d_{modal,bed}$), 10$^{th}$ percentile ($d_{10,bed}$), and 90$^{th}$ percentile ($d_{90,bed}$) grain diameters of bed sediments at each field site, as well as the impact threshold shear stresses ($\tau_{it}$) and median airborne grain diameters ($d_{50,air}$) at these sites. Uncertainties on reference grain sizes denote standard errors computed from variation among individual samples. Impact thresholds $\tau_{it}$ and their uncertainties were computed by Martin and Kok (2018). As described in Sec. 2, Oceano measurements are separated into distinctive time periods (1 and 2) and spatial clusters (N and S). The error ranges presented here reflect only measurement uncertainties and may not fully account for systematic uncertainties in the sampling protocols or analysis methods.

| Site | | $d_{50,bed}$ (mm) | $\bar{d}_{bed}$ (mm) | $d_{modal,bed}$ (mm) | $d_{10,bed}$ (mm) | $d_{90,bed}$ (mm) | $\tau_{it}$ (Pa) | $d_{50,air}$ (mm) |
|---|---|---|---|---|---|---|---|---|
| **Jericoacoara** | | 0.526 ±0.011 | 0.422 ±0.034 | 0.603 ±0.039 | 0.097 ±0.004 | 0.847 ±0.011 | 0.111 ±0.002 | 0.332 ±0.040 |
| **Rancho Guadalupe** | | 0.533 ±0.009 | 0.471 ±0.033 | 0.603 ±0.000 | 0.219 ±0.012 | 0.839 ±0.012 | 0.110 ±0.002 | 0.387 ±0.013 |
| **Oceano (May 15-22, 2015)** | N1 | 0.332 ±0.007 | 0.321 ±0.028 | 0.349 ±0.065 | 0.181 ±0.002 | 0.551 ±0.010 | 0.083 ±0.001 | 0.228 ±0.069 |
| | S1 | 0.397 ±0.011 | 0.362 ±0.038 | 0.516 ±0.065 | 0.185 ±0.005 | 0.625 ±0.010 | | 0.248 ±0.010 |
| **Oceano (May 23-June 4, 2015)** | N1 | 0.406 ±0.007 | 0.376 ±0.043 | 0.516 ±0.078 | 0.196 ±0.003 | 0.650 ±0.007 | 0.083 ±0.001 | 0.235 ±0.013 |
| | S2 | 0.472 ±0.010 | 0.415 ±0.056 | 0.558 ±0.084 | 0.199 ±0.007 | 0.713 ±0.007 | | 0.250 ±0.045 |

## 2. Methods

To evaluate the size-selectivity of susceptibility and mobility in aeolian saltation (Eq. 2), we analyzed simultaneous measurements of saltation and wind at three field sites (Fig. 2): Jericoacoara, Ceará, Brazil (2.7969°S, 40.4837°W); Rancho Guadalupe, California, U.S.A. (34.9592°N, 120.6431°W); and Oceano, California, U.S.A. (35.0287°N, 120.6277°W). All field sites contain mostly flat, unvegetated, sand-covered surfaces, with distinctive bed sediment PSDs. To characterize height-specific saltation fluxes $q(z)$ and airborne PSDs, Big Springs Number Eight (BSNE) saltation traps (Fryrear, 1986) collected saltating sand particles over ~1 hour intervals at multiple heights ($z = 0.05 - 0.7$m) above the sediment bed (see Martin et al., 2018 for further detailed description of field sites and saltation measurement methods). A sonic anemometer at ~0.5 m measured wind velocities over time intervals corresponding to the BSNE trap collections; from these data, we calculated wind shear stresses $\tau$ by the Reynolds stress method (see Martin and Kok, 2017 for further detailed description of wind data processing methods).

### 2.1. Bed surface grain size characteristics

To characterize grain size characteristics of bed surface particles at each field site, we collected daily grab samples from the top ~1 cm of the bed surface. We then ran each of these surface grab samples through a Camsizer particle size analyzer (Jerolmack et al., 2011) to obtain particle size distributions $\frac{dV}{d\ln(d)}$, where $dV$ is the volume fraction of particles in the $d$ grain diameter size bin, non-dimensionalized by the logarithm of the size range of the bin $d\ln(d)$. For each site, we then obtain the characteristic bed surface grain-size distributions $\frac{dV_{bed}}{d\ln(d)}$ by computing averages of $\frac{dV}{d\ln(d)}$ for surface grab samples collected at each field site. Based on each $\frac{dV_{bed}}{d\ln(d)}$, we calculate reference grain diameters $d_{10,bed}$, $d_{50,bed}$, and $d_{90,bed}$, from the respective 10th, 50th, and 90th percentile values of the PSD, we calculate mean grain diameter $\bar{d}_{bed}$ as the geometric mean of the PSD, and we calculate modal grain size $d_{bed,modal}$ as the particle size bin with the largest non-dimensionalized volume fraction in the PSD (Table 1).

Measurements collected at Oceano were more spatially and temporally extensive than at Jericoacoara and Rancho Guadalupe (Fig. 2). In particular, we deployed BSNE traps at Oceano in two distinctive clusters separated in the spanwise direction by about 8 meters, and we noticed significant differences in bed PSDs between these two clusters. Thus, we henceforth analyze clusters of BSNE measurements from the north and south side of the Oceano field site as if they were from two separate field sites; we denote these as "Oceano N" and "Oceano S." (Because the anemometer at Oceano was positioned halfway between both BSNE clusters, we assign identical $\tau$ and $u_*$ values to concurrent time intervals for Oceano N and S.) Furthermore, we noticed a significant coarsening of the bed PSD over time at Oceano (see Supporting Information Fig. S1); thus, we further subdivide data roughly into the first (May 15-22, 2015) and second (May 23-June 4, 2015) halves of the deployments, resulting in a total of four "virtual" field sites at Oceano: Oceano N1 and S1 for the first half, and Oceano N2 and S2 for the second half. Spanwise separation among BSNEs analyzed together at Jericoacoara, Rancho Guadalupe, and within each Oceano cluster was always less than 4m (Fig. 2), and each set of cluster samples showed no systematic spatial variation among PSDs. Unless otherwise noted, we henceforth refer to virtual field sites simply as "sites" for simplicity. Thus, we have a total of 6 sites for analysis.

Site-averaged bed surface PSDs are displayed in Fig. 3, and reference grain sizes for these sites are listed in Table 1. Each site displays a distinctive bed PSD. Both Jericoacoara and Rancho Guadalupe have a similar median bed surface particle diameter ($d_{50,bed} \approx 0.53$ mm), but Jericoacoara displays a second fine mode (0.11 mm) that is lacking elsewhere. PSDs are generally finer at the Oceano sites than at Jericoacoara and Rancho Guadalupe. At Oceano, median bed surface particle diameters are finest on the north side of the field site and in the first half of the deployment ($d_{50,bed} \approx 0.33$ mm, 0.40 mm, 0.41 mm, and 0.47 mm for Oceano N1, S1, N2, and S2, respectively). All bed PSDs are slightly skewed toward finer sizes; thus, at all sites, modal grain size $d_{modal,bed}$ exceeds median grain size $d_{50,bed}$, which in turn exceeds mean grain size $\bar{d}_{bed}$ (Table 1).

## 2.2. Airborne grain size characteristics

We obtained airborne PSDs by performing Camsizer analyses on sediment samples collected in BSNE traps. For certain periods of low saltation flux, some BSNE traps at heights far above the bed surface collected insufficient sediment (i.e., ≲10 g) for grain-size analyses. In these cases, we combined samples from adjacent time intervals for the same BSNE height prior to running Camsizer grain-size analyses. Furthermore, because BSNE trap efficiency is known to degrade for $d \lesssim 0.13$ mm (Goossens et al., 2000), we limit analyses involving airborne PSDs to size classes for which $d \geq 0.13$ mm.

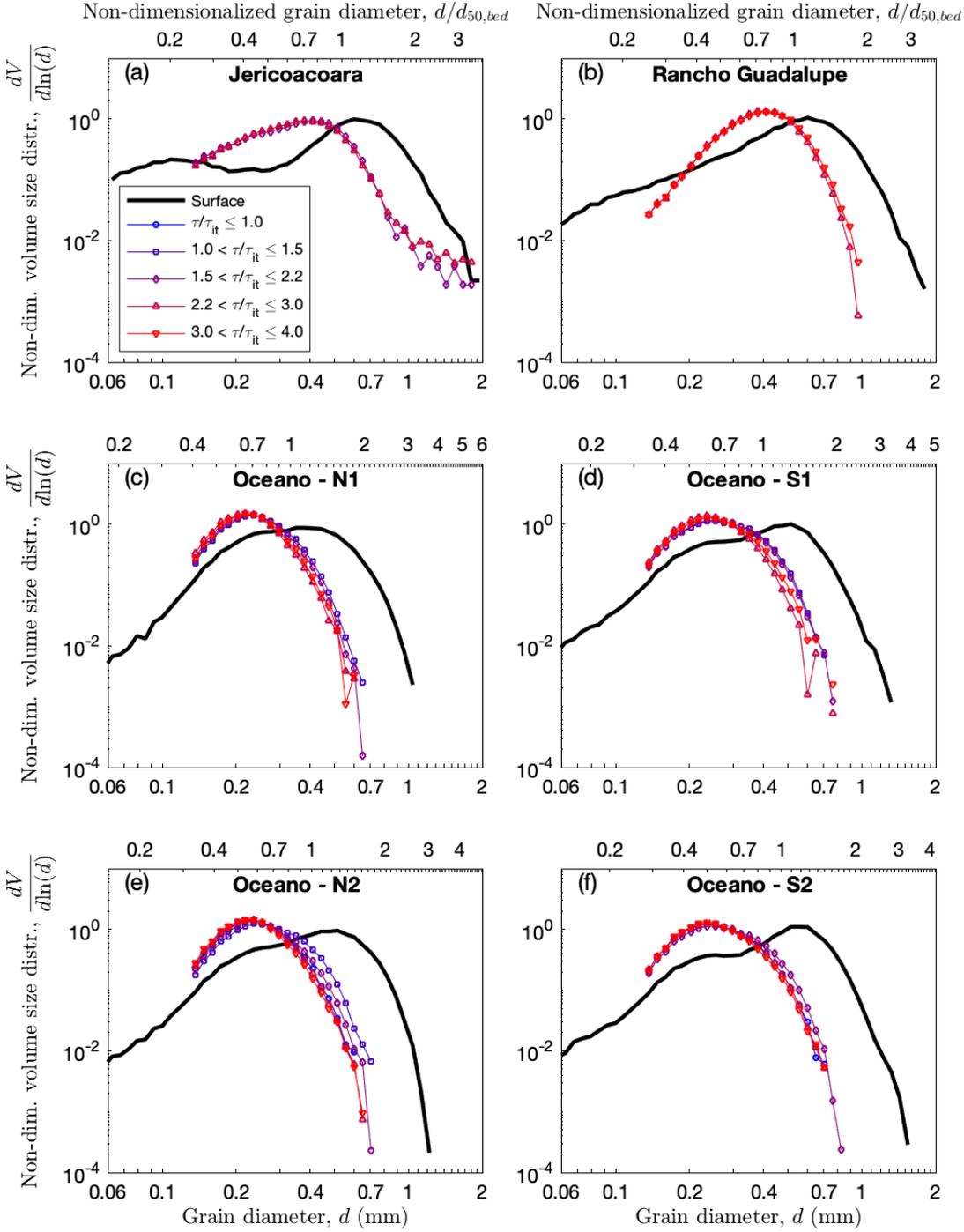

**Figure 3.** Site-averaged bed surface particle size distributions (PSDs), $\frac{dV_{bed}}{d\ln(d)}$, and airborne PSDs, $\frac{dV_{air}}{d\ln(d)}$. The airborne PSDs are conditioned on specific ranges of shear stress $\tau$ and non-dimensionalized by impact threshold shear stress $\tau_{it}$ for each site. Results show that a wide range of size classes move whenever saltation is active, implying equal susceptibility to saltation transport for those size classes. Airborne PSDs are limited to $d \geq 0.13$ mm to account for BSNE trap inefficiency for small grain sizes (Goossens et al., 2000).

We calculate a vertically-integrated airborne PSD $\frac{dV_{air}}{d\ln(d)}$ for each ~1 hr measurement interval as a weighted sum of the individual PSDs from the vertical profile of BSNEs contained within this time interval. As we describe in detail in Appendix A, each measured PSD is weighted both by its vertical coverage and by its contribution to the vertically-integrated saltation flux. In performing this weighting, we assume exponential saltation flux profiles of the form $q(z) = q_0\exp(-z/z_q)$ (Martin et al., 2018), where $q$ is the BSNE-measured height-specific saltation flux at height $z$, and $q_0$ and $z_q$ are fitting parameters. Plots of individual saltation profiles (see Supporting Information Fig. S2) confirm the exponential nature of these profiles.

Because of the large size of BSNE traps, the airborne PSDs measured at our field sites were mostly obtained relatively far above the bed surface. At Jericoacoara and Rancho Guadalupe, this minimum trap airborne PSD height was roughly $z = 10$ cm; at Oceano, where we deployed a "modified BSNE" closer to the surface, the minimum height was roughly $z = 5$ cm (Martin et al., 2018). Expressed in terms of $z_q$, the characteristic "$e$-folding" saltation height for the each profile, most PSD measurements covered only $z/z_q \geq 1$ (see Supporting Information Fig. S3). In other words, our calculated values for $\frac{dV_{air}}{d\ln(d)}$ represent vertically-integrated airborne PSDs only for the upper portions of saltation flux profiles. In Sec. 3.2, we estimate how this limited range of measurements affects our interpretations of size-selective mobility, and we further address these limitations in Discussion Sec. 4.1.

## 3. Results

In this section, we examine the shear stress dependence of airborne PSDs to show that most particle size classes in aeolian saltation display equal susceptibility; that is, the threshold wind stress for transport is independent of particle size (Sec. 3.1). Noting the limitations of our measurements, which do not include near-surface saltation PSDs, we also briefly analyze how the relative mobilities of individual size classes, namely the ratios of airborne and surface PSDs, vary with particle size (Sec. 3.2).

### 3.1. Equal susceptibility in aeolian saltation

To determine whether saltation susceptibility varies with particle size, we examine the variation in airborne PSDs with wind shear stress $\tau$. At each site, we group vertically-integrated airborne grain-size distributions $\frac{dV_{air}}{d\ln(d)}$ for given ranges of the concurrently-measured shear stress ratios $\tau/\tau_{it}$. Here, $\tau_{it}$ is the bulk (non-size-selective) impact threshold shear stress, which describes the minimum wind stress for sustaining saltation, and which we determined independently for each site (Martin and Kok, 2018). Fig. 3 shows that, at each site, the airborne PSD is substantially finer than the surface PSD. Furthermore, $\tau/\tau_{it}$-conditioned airborne PSDs are mostly insensitive to variation in $\tau/\tau_{it}$, with a slight fining of the airborne PSD occurring only at the Oceano sites. For a substantial range of particle size classes ($d \lesssim 1$ mm at Jericoacoara and Rancho Guadalupe, $d \lesssim 0.6$ mm at the four Oceano sites), all particle sizes are mobilized whenever saltation is active, regardless of the exact value of $\tau/\tau_{it}$.

Based on the relative insensitivity of airborne PSDs to variation in shear stress displayed in Fig. 3, we infer that equal susceptibility occurs for non-dimensionalized grain sizes $d/d_{50,bed} \lesssim 1.5$ at all sites, and thus that transport occurrence for this wide range of particle sizes is governed by the single bulk threshold $\tau_{it}$ (or equivalent impact threshold shear velocity $u_{*,it}$),. For $d/d_{50,bed} \gtrsim 1.5$, airborne PSDs display a finite upper limit grain size at all sites except Jericoacoara. This upper limit could indicate that there is a fixed maximum grain size susceptible to saltation at each site, or it could simply reflect detection limits in measurements of PSDs. For $d \lesssim 0.13$ mm, our ability to interpret particle susceptibility is limited by the reliability of the BSNEs (Goossens et al., 2000). Despite this reliability issue, all airborne PSDs show the presence of all particle size classes down to $d \sim 0.07$ mm (roughly, the lower limit grain size for transition from saltation to full suspension (Kok et al., 2012)), indicating that there is no clear lower size limit on equal susceptibility.

In contrast to the other sites, $\frac{dV_{air}}{d\ln(d)}$ for Jericoacoara displays a very coarse tail (i.e., $d \gtrsim 1$ mm). Examining the individual airborne PSDs constituting the averaged $\frac{dV_{air}}{d\ln(d)}$ for Jericoacoara, we find that $d > 1$ mm particles are contained in many of these individual PSDs, but only for those airborne samples collected closer to the surface (i.e., $z/z_q < 2$). This could indicate that coarse reptating particles, i.e. those particles that are ejected but do not experience subsequent rebound (e.g., Bauer and Davidson-Arnott, 2014), are occasionally lofted into the middle of the saltation layer, or it could simply suggest the presence of wind-blown coarse debris at this site.

### 3.2. Size-selective mobility in aeolian saltation
We now use our finding of equal susceptibility for non-dimensionalized grain sizes $d/d_{50,bed} \lesssim 1.5$ (Sec. 3.1) to explore the size-selectivity of saltation mobility. We do so by examining the relative contributions of different particle size classes $i$ to the bed surface and airborne PSDs. Equal susceptibility implies that $\tau_{th,i} = \tau_{it}$ for all $i$ in Eq. 2 (see Fig. 1). Defining the size-selective saltation flux as $Q_i = Q f_{air,i}$, where $Q$ is the bulk (non-size-selective) saltation flux, and $f_{air,i}$ is the fractional contribution of size class $i$ to the total PSD, we can simplify Eq. 2 to directly relate the size-selective mobility parameter $C_i$ to the ratio of site-averaged airborne fraction $\langle f_{air,i} \rangle$ and bed fraction $f_{bed,i}$ at each site:
$$\frac{C_i}{C} = \frac{\langle f_{air,i} \rangle}{f_{bed,i}}, \qquad (3)$$
where $C$ is the bulk flux law coefficient. By computing the average airborne fraction $\langle f_{air,i} \rangle$, we are neglecting variations in $f_{air,i}$ with $\tau$, which we have shown to be small (Fig. 3). Eq. 3 allows us to directly relate measured ratios of airborne to soil volume fractions, $\langle f_{air,i} \rangle / f_{bed,i}$, to size-selective mobilities $C_i$, such that $\langle f_{air,i} \rangle / f_{bed,i} > 1$ (i.e., $C_i/C > 1$) indicates enhanced mobility, and $\langle f_{air,i} \rangle / f_{bed,i} < 1$ (i.e., $C_i/C < 1$) indicates reduced mobility.

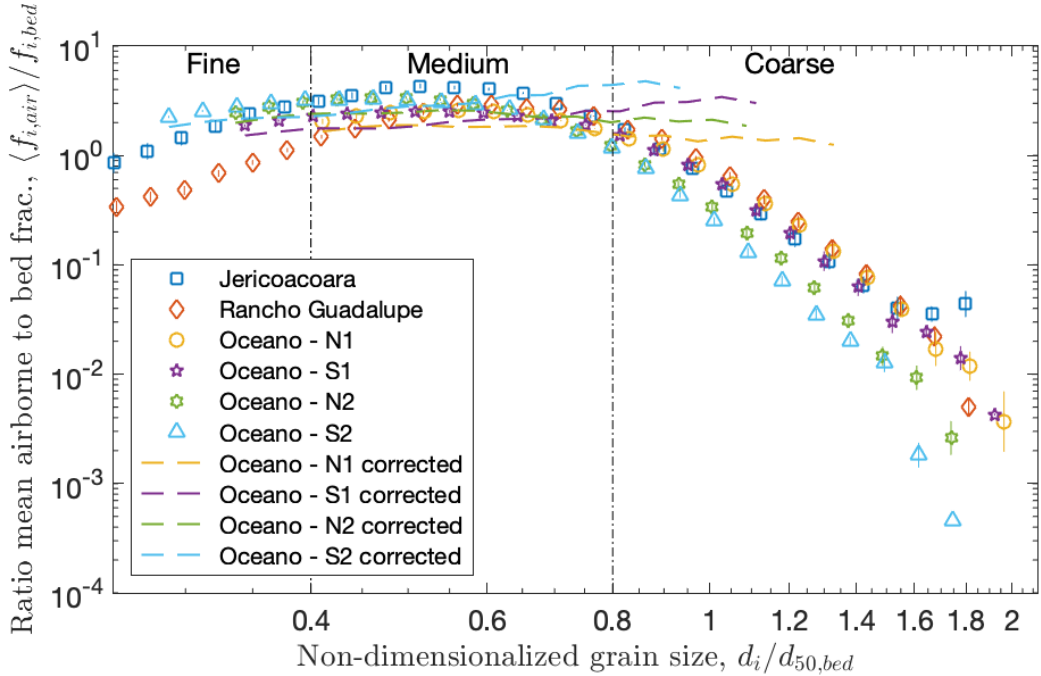

**Figure 4.** Size-selective mobility at each site, as denoted by the ratio of mean airborne and bed surface particle volume fractions, $\langle f_{air,i}\rangle/f_{bed,i}$, versus non-dimensionalized particle size $d_i/d_{50,bed}$. Dashed lines indicate boundaries between designated ranges for fine, medium, and coarse particles. Volume fraction ratios are shown only for $d \geq 0.13$ mm (BSNE trap reliability limit) and for $d \leq 1$ mm (maximum airborne grain size at all sites except Jericoacoara). Error bars indicate standard errors computed from the individual $f_{air,i}$ values in each $\langle f_{air,i}\rangle$. Though suggestive of size-selectivity in aeolian saltation mobility, our measurements only account for particles far above the bed surface (i.e., $z/z_q \gtrsim 1$). To estimate the effect of this limitation, we perform a "correction" (shown as the dashed lines) for our Oceano measurements, based on correction factors computed from the saltation PSD profiles measured Namikas (2006) (see Fig. 5 below). With this correction applied, size-selectivity in particle mobility is mostly eliminated.

Fig. 4 shows how the $\langle f_{air,i}\rangle/f_{bed,i}$ ratio, and thus $C_i/C$, varies with non-dimensionalized particle size class $d_i/d_{50,bed}$ at each site. All sites show a similar pattern of substantial selective mobility of different particle sizes $d_i$. In general, we can subdivide size-selective mobility trends into three non-dimensionalized particle size classes. "Medium" particles, which we define as covering the approximate range $0.4 \lesssim d_i/d_{50,bed} \lesssim 0.8$, display the highest mobility, which is enhanced in comparison to the average particle mobility. "Fine" particles, which we define as $d_i/d_{50,bed} \lesssim 0.4$, experience average or somewhat reduced mobility, which decreases slightly with decreasing $d_i/d_{50,bed}$. "Coarse" particles, which we define as $d_i/d_{50,bed} \gtrsim 0.8$, display a rapid reduction in mobility with increasing $d_i/d_{50,bed}$.

Though our results are suggestive of size-selectivity in particle mobility, a lack of near-surface saltation measurements casts doubt on the universality of these results. To estimate the effect of missing near-surface measurements, we derive and apply a "correction factor" based on the measurements of Namikas (2006). We reanalyze the height- and particle size-resolved horizontal sand flux data reported in Figure 3 of Namikas (2006) to calculate both far-from-surface PSDs (i.e., $z/z_q \gtrsim 1$), as in our study, and airborne PSDs over most of the height range (i.e., for $z/z_q \gtrsim 0.1$), beyond the limits of our study. Because the Namikas study was performed very close to our Oceano sites, we assume that its measured vertical PSD patterns are similar to those at our site.

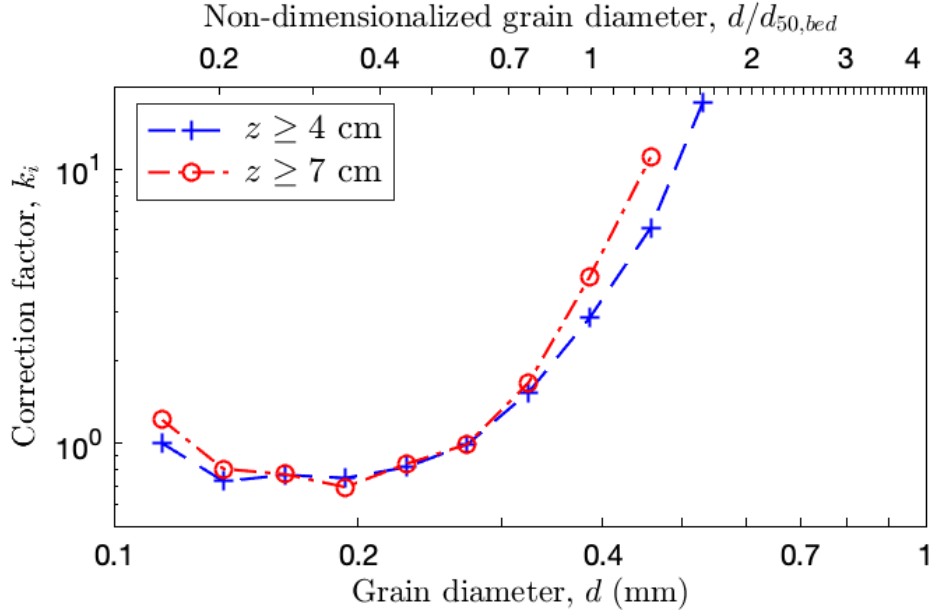

**Figure 5.** Airborne PSD correction factors $k_i$ derived from the measurements of Namikas (2006), as a function of grain diameter $d$. The value of $k_i$ (Eq. 4) estimates the ratio of the actual airborne volume fraction for $d$ versus the value obtained with PSD measurements only at a height of $z \geq 4$ cm (blue line) and $z \geq 7$ cm (red line). $k_i < 1$ indicates an overestimate of the actual airborne volume fraction for $d$; $k_i > 1$ indicates an underestimate for $d$.

We derive a size-resolved correction factor $k_i$ from the Namikas measurements as:
$$k_i = \frac{f_{air,i}}{f_{air,i}|_{z/z_q \geq 1}} \quad (4)$$
where $f_{air,i}$ refers to the airborne particle volume fraction computed over the full range of $z$, and $f_{air,i}|_{z/z_q \geq 1}$ refers to the airborne particle volume fraction only for $z/z_q \geq 1$ (the approximate height range of our PSD measurements). We then compute "corrected" mean airborne volume fractions $\langle f_{air,i} \rangle_{corrected}$ at Oceano as:
$$\langle f_{air,i} \rangle_{corrected} = k_i \langle f_{air,i} \rangle. \quad (5)$$
We choose a cutoff height $z$ for computation of the correction factor (Eq. 4) based on the mean saltation layer height measured at Oceano, which was $z_q = 5.5$ cm (Martin and Kok, 2017). However, the closest heights of PSD measurements by Namikas were at $z = 4$ cm and $z = 7$ cm. We thus choose to compute $k_i$ both for $z \geq 4$ cm and $z \geq 7$ cm. The resulting correction factors are shown in Fig. 5.

The effect of applying the correction factors (we use the more conservative $k_i$ from the $z \geq 7$ cm case in Fig. 5) is shown by the dashed lines in Fig. 4. For the range of particle sizes captured in the Namikas (2006) measurements, the correction mostly eliminates any apparent size-selectivity in the volume fractions. We thus conclude that our study misses a large component of coarse saltating particles close to the surface and below the e-folding height $z_q$, and therefore that we cannot generalize from the apparent size-selectivity shown in Fig. 4 to draw any definitive conclusions about size-selective mobility in aeolian transport. We further explore the causes and effects of missing near-surface measurements in Sec. 4.1 below.

## 4. Discussion

Our analyses indicate that equal susceptibility occurs in aeolian saltation for all particle size classes up to ~1.5 times the median diameter of particles on the surface of the soil bed (Fig. 3). In contrast, it is not clear from our measurements whether particle mobility is size-selective (Figs. 4 and 5).

In this Discussion section, we explore the physical meaning and implications of observed size-resolved patterns in aeolian saltation. To do this, we first examine vertical variation in PSDs and the effect of a limited sampling height range on our inferences of size-selective mobility and equal susceptibility (Sec. 4.1). Then, we investigate two other potential limitations of our analyses (Sec. 4.2). Next, we consider the physical meaning of equal susceptibility in aeolian saltation (Sec. 4.3). Finally, we explore the implications of equal susceptibility for the modeling of saltation-driven dust emission and aeolian geomorphic processes (Sec. 4.4).

### 4.1. Vertical variation in PSDs and the effect of a limited sampling height range

As described in Sec. 3.2, a major limitation of our analyses is that measured airborne PSDs capture only dynamics in the upper part of the saltation profile, at heights exceeding the *e*-folding saltation layer height $z_q$. When a correction (Fig. 5) is applied to account for near-surface saltators on overall airborne PSDs, size-selective mobility is eliminated. This likely reflects the fact that coarse particles are disproportionately represented near the surface, whereas fine particles are disproportionately represented farther from the surface. This vertical size fractionation is supported by the field measurements of Namikas (2006) as well as by Farrell et al. (2012), who also showed a more broadly distributed (i.e., poorly sorted) distribution of airborne particle sizes near the bed surface.

To further understand the effect of measurement height on airborne PSDs, we plot vertical profiles of median airborne particle diameters $d_{50,air}$ from Namikas (2006), Farrell et al. (2012), and at our at our sites (Fig. 6). For the sake of comparison, we non-dimensionalize $d_{50,air}$ by median surface diameter $d_{50,bed}$ at each site, and we non-dimensionalize height above the bed surface $z$ by mean *e*-folding saltation layer height $\langle z_q \rangle$. We further subdivide measurements by shear stress $\tau$ (non-dimensionalized by impact threshold stress $\tau_{it}$) to reveal any possible shear stress dependence.

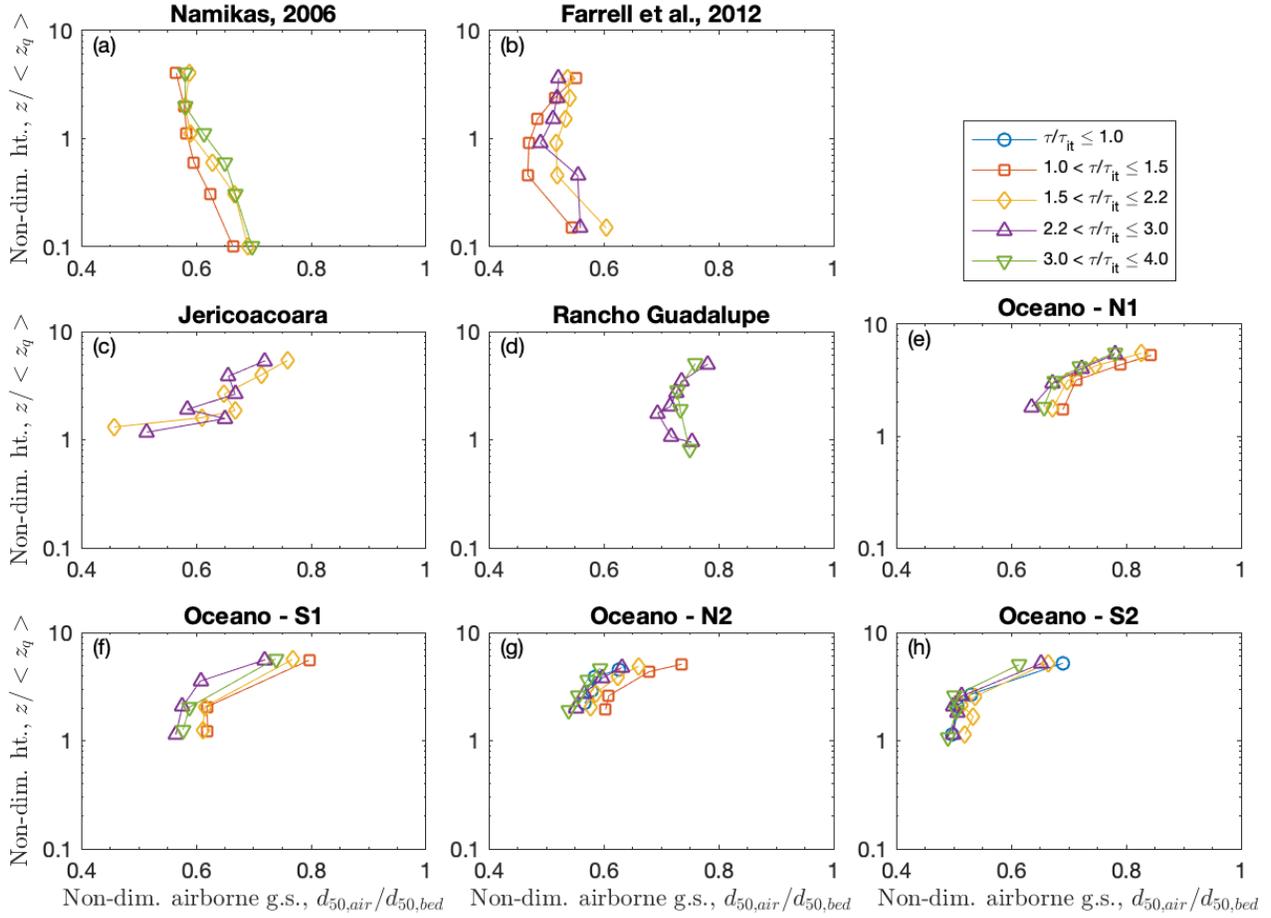

**Figure 6.** Median airborne grain diameter $d_{50,air}$ versus height above the bed surface $z$, for measurements (a) by Namikas (2006), (b) by Farrell et al. (2012), and (c-h) at our field sites. $d_{50,air}$ is non-dimensionalized by median bed grain diameter $d_{50,bed}$, and $z$ is non-dimensionalized by mean $e$-folding saltation layer height $\langle z_q \rangle$. Measurements are further subdivided by shear stress $\tau$, non-dimensionalized by impact threshold stress $\tau_{it}$. Namikas and Farrell et al. did not report $\tau_{it}$ or bed PSDs; thus, we use the $\tau_{it}$ and mean $d_{50,bed}$ from our Oceano site for Namikas and from our Jericoacoara site for Farrell et al. (see Martin and Kok, 2017 for values), based on the close proximity of their studies to our field sites. Farrell et al. further did not report full airborne PSDs or $d_{50,air}$, thus we assume $d_{50,air}$ to equal the mean airborne grain diameters reported in their study. Values for $z_q$ for the Namikas and Farrell et al. studies were calculated in Martin and Kok (2017).

### 4.1.1. Effects of limited sampling height range on inferred size-selective mobility

The Namikas (2006) and Farrell et al. (2012) measurements in Fig. 6 reinforce the importance of near-surface coarsening of the airborne PSD, which is missing from our measurements. (We note similar observations for $10^{th}$ percentile and $90^{th}$ percentile particle sizes – see Supporting Information Figs. S4 and S5). Such underestimation of the coarse contribution may be exacerbated by the fact that saltation flux is greatest near the surface, perhaps even exceeding the expectation of the near-surface flux from fitting an exponential profile (e.g., Namikas, 2003; Ni et al., 2003; Creyssels et al., 2009; Bauer and Davidson-Arnott, 2014). The concurrent coarsening of near-surface PSD and overall bulge in the near-surface flux profile likely represent the contribution of reptation – the low-energy non-sustaining particle trajectories that result from the splash of surface particles (e.g., Durán et al., 2011).

In addition to near-surface coarsening, Fig. 6 shows that airborne particle sizes become coarser far above the bed surface, both in the measurements of Farrell et al. (2012) and at all of our field sites. This far-from-surface coarsening suggests that, though fewer coarse particles are ejected into full saltation trajectories, those ejected coarse particles that do become saltators (instead of low energy reptators) achieve taller trajectories than their fine particle companions. However, because of the exponential nature of the overall vertical profile of saltation, these far-from-surface coarse saltators (which are included in our measurements), contribute far less to overall airborne PSDs than near-surface coarse saltators (which are not included in our measurements). Given the limitations of our study, we suggest that future studies be conducted to examine the effects of vertical fractionation of particle sizes on possible size-selectivity in particle mobility.

### 4.1.2. Effects of limited sampling height range on inferred equal susceptibility

Despite the above limitations on our interpretations of size-selective mobility, we argue here that the limited vertical range of our PSD measurements does not affect our overall inference about the occurrence of equal susceptibility. Regardless of whether or not near-surface measurements are considered, all vertical PSD profiles display a similarly small amount of variation in $d_{50,air}$ with shear stress $\tau$ (Fig. 6). The same is generally true for $d_{10,air}$ and $d_{90,air}$ (Figs. S4 and S5). Furthermore, regardless of whether or not near-surface observations are included, airborne PSDs measured by Namikas (2006) show the presence of all particle size classes for all values of $\tau$ (Fig. 7), in a manner similar to the observations at our field sites (Fig. 3).

One factor that is affected by the inclusion or exclusion of near-surface measurements is the specific manner in which airborne PSDs vary with $\tau$. For the measurements of Namikas (2006), inclusion of all measurement heights produces a slight coarsening of the overall airborne PSD with increasing $\tau$ (Fig. 7a), whereas inclusion of only measurements for $z \geq 7$ cm (Fig. 7b) eliminates any appreciable variation in airborne PSDs with shear stress. This behavior of the far-from-surface PSDs measured by Namikas resembles the negligible stress-dependence in airborne PSDs at our sites (Fig. 3). This difference in stress-dependence of PSDs with the range of measurement heights likely reflects the presence or absence of the effect of near-surface coarse saltators. Whereas Fig. 6 shows that the upper parts of the profiles tend to show no change or even a slight fining with increasing $\tau$, the lower parts of the profiles measured by Namikas and Farrell et al. show an appreciable coarsening with increasing $\tau$. Thus, as we explain further in Sec. 4.4.2 below, it does remain possible that very coarse particles, such as those associated with armoring on granule ripples, do experience selective susceptibility due to the distinctive mechanisms for mobilizing these particles.

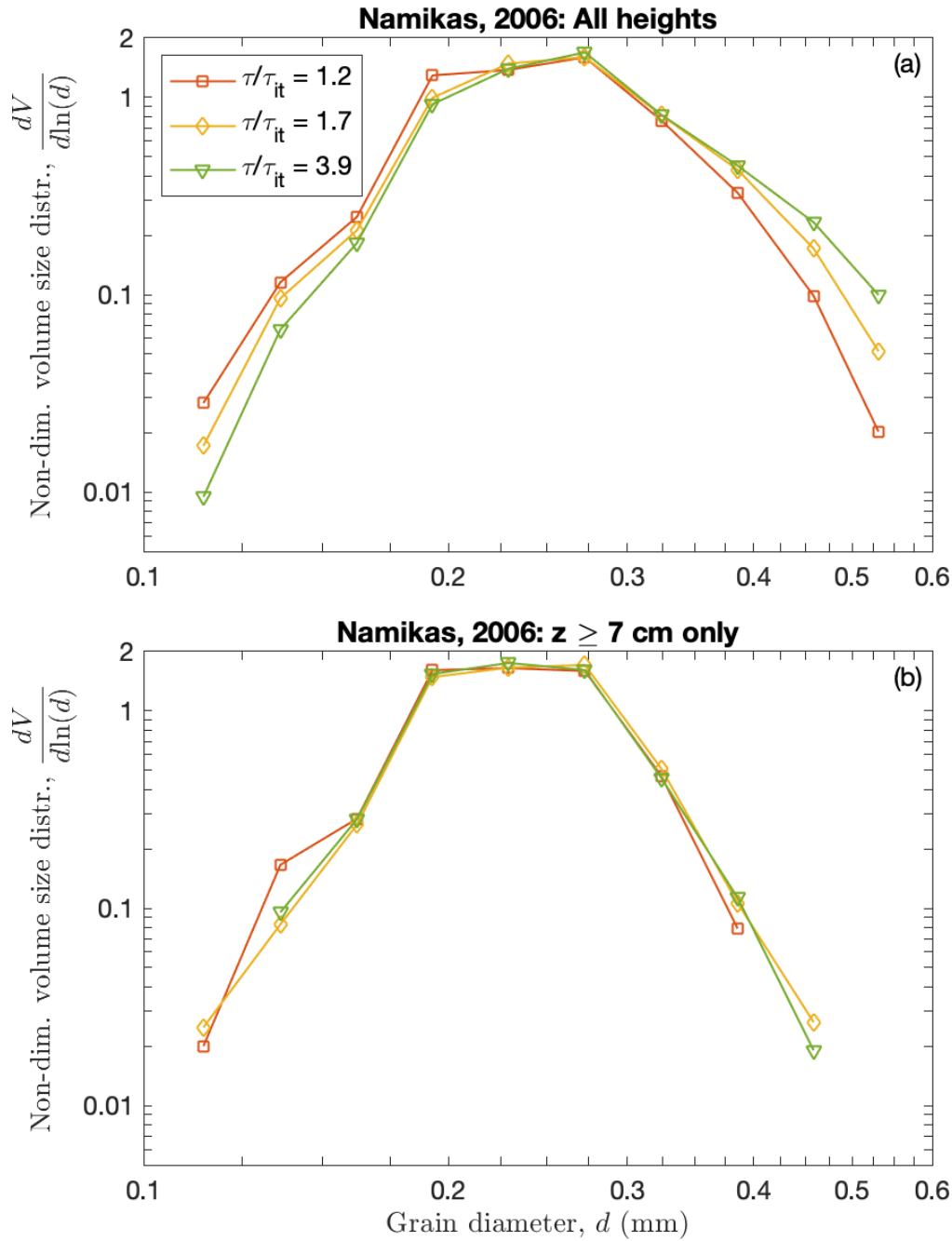

**Figure 7.** Shear stress-dependent airborne PSDs measured by Namikas (2006) for (a) full vertical profiles and (b) only $z \geq 7$ cm. Stresses are non-dimensionalized by impact threshold shear stress $\tau_{it}$. Namikas did not report $\tau_{it}$; thus, we use the $\tau_{it}$ from our Oceano site (see Martin and Kok, 2017), based on the close proximity of his study to our field site.

### 4.2. Additional limitations of saltation susceptibility and mobility analyses

One additional limitation of our analyses is that our observations are confined to a relatively small sampling of field sites, all of which contain relatively coarse and widely mixed (i.e., poorly sorted) grain sizes. In contrast, sand on desert dunes typically contains finer and more uniform (i.e., well sorted) grain sizes (e.g., Lancaster, 1981, 1986). In wind tunnel experiments containing medium ($d_{modal,bed} \approx 0.4$ mm), highly mixed sediment beds, Williams (1964) observed substantial fining of airborne PSDs with respect to surface PSDs. In field measurements of saltation over finer ($d_{modal,bed} \approx 0.2$ mm), less mixed sediment beds, Gillette and Walker (1977) observed only a slight fining of airborne PSDs. These somewhat conflicting reports on how airborne PSDs vary with soil properties suggest the need for further work to understand how size-selective mobility depends on the PSD of the soil bed.

A second additional limitation of our analyses is that our measurements poorly represent the dynamics of fine particles. Due to BSNE collection efficiency issues for $d \lesssim 0.13$ mm particles (Goossens et al., 2000), we chose to exclude these fine particles from our analyses of airborne samples. At Oceano N1, this limited our analyses to particles in the medium and coarse ranges, i.e., $d_i/d_{50,bed} \gtrsim 0.4$. At the Jericoacoara and Rancho Guadalupe sites, where a coarser median bed particle diameter allowed for analysis of particles as small as $d_i/d_{50,bed} \approx 0.25$, fine particles appear to have experienced a significant reduction in relative volume fraction (Fig. 4) with decreasing $d_i/d_{50,bed}$. However, along with the sampling height issues already mentioned, we lack the data at Oceano to fully evaluate these possible volume fraction patterns for fine particles.

### 4.3. Physical meaning of equal susceptibility

Our observations strongly support the existence of equal susceptibility for a wide range of grain-size classes, which we infer from the presence of all $d_i/d_{50,bed} \lesssim 1.5$ particle sizes in the saltation layer, regardless of the specific shear stress $\tau$ driving saltation (Fig. 3). Even in cases of weak, intermittent saltation (i.e., $\tau/\tau_{it} \leq 1$ at Oceano N2 and S2, Figs. 3e and 3f), all $d_i/d_{50,bed} \lesssim 1.5$ particle size classes are present in the saltation layer. Though we cannot rule out the possibility that $d_i/d_{50,bed} \gtrsim 1.5$ particles were saltating near the surface (i.e., $z/\langle z_q \rangle < 1$) where we did not make PSD measurements (see Sec. 4.1 above), our comparison to other studies (Figs. 6 and 7) indicates that equal susceptibility also holds when considering this lower part of the saltation layer. Furthermore, the fact that the coarsest saltators were observed far above the bed suggests that all of these size classes were participating in energetic saltation trajectories, not just reptation and low-energy saltation. With the exception of Jericoacoara, there is also a relatively abrupt cutoff in particle susceptibility for $d_i/d_{50,bed} \gtrsim 1.5$ particles, even at larger shear stresses.

The combined occurrence of equal susceptibility for $d_i/d_{50,bed} \lesssim 1.5$ and an equal lack of susceptibility for $d_i/d_{50,bed} \gtrsim 1.5$, regardless of the value of $\tau$, supports models for aeolian saltation in which particle entrainment is dominantly governed by splash (e.g., Anderson and Haff, 1988), and in which the trajectories of particles impacting with the bed are mostly unaffected by the wind shear stress $\tau$ (Andreotti, 2004; Durán et al., 2011; e.g., Ungar and Haff, 1987). Such models therefore predict that impact energies, and thus also ejection rates, mostly remain constant with shear stress $\tau$ (Rasmussen and Sørensen, 2008; Creyssels et al., 2009; Ho et al., 2011; Kok et al., 2012). Recent model simulations by Durán et al. (2014) further indicate that the saltation layer is separated from the static sediment bed by a mobile "interfacial" layer, a few grain diameters thick, which is dominated by mid-air collisions among particles. Thus, particle susceptibilities to saltation are tied to the dynamics of this layer rather than the static configuration of the bed. Though these model predictions were obtained primarily for beds of uniformly-sized particles, the observation of wind-invariant mean saltation trajectories at field sites containing natural mixed sediment beds (Greeley et al., 1996; Namikas, 2003; Farrell et al., 2012; Martin and Kok, 2017) indicates that wind-invariant impact speeds also occur in mixed-bed settings. Therefore, regardless of wind strength, bed particle susceptibility to saltation is governed by a mostly constant distribution of impact speeds, which in turn determines the fixed upper limit particle size susceptible to saltation (i.e., $d_i/d_{50,bed} \approx 1.5$ in our study). Particles larger than this upper limit size may still be ejected by particle impacts, but such ejecta fail to become saltators, instead following small creep or reptation trajectories (e.g., Andreotti, 2004). It remains possible that, unlike for saltation, the size-selective susceptibility to reptation does vary with grain size (Isenberg et al., 2011), because the high-energy tail of the impact speed distribution broadens with increasing shear stress (Ho et al., 2014; Kok et al., 2014). Such departures from equal susceptibility due to reptation may be visible in the shear stress-dependence of the coarse tail of the airborne PSD for near-surface saltators (i.e., Figs. 6 and 7).

### 4.4. Implications for modeling aeolian geomorphology and dust emission

Developing an accurate understanding of size-selective saltation fluxes, and how these depend on the soil PSD, is critical both for modeling aeolian geomorphic and stratigraphic processes driven by saltation (e.g., Anderson and Bunas, 1993; Lindhorst and Betzler, 2016) and for determining the saltation-driven emission rate for airborne dust particles (e.g., Marticorena and Bergametti, 1995; Shao, 2008; Kok et al., 2014). In this section, we evaluate existing approaches to modeling aeolian saltation over natural mixed soils. Based on this evaluation, we then offer suggestions for how to properly represent equal susceptibility and size-selective mobility in the study of aeolian processes on Earth and planetary surfaces.

### 4.4.1. Evaluation of existing treatments of saltation over mixed sediment beds

As noted in the Introduction, there are two primary existing approaches to modeling saltation over mixed sediment beds: First, the 'independent saltation' approach, in which each particle size class is modeled independently as if it were the only particle size on the bed (Marticorena and Bergametti, 1995; Shao et al., 1996; Alfaro and Gomes, 2001; Zender et al., 2003), and second, the 'representative saltation' approach, in which a single reference grain size is chosen to represent the saltation dynamics for all particle sizes (Elbelrhiti et al., 2005; Claudin and Andreotti, 2006; Andreotti et al., 2010). We consider these two approaches here.

Our finding of equal susceptibility refutes the independent saltation approach to determining saltation thresholds, at least for a reasonably wide distribution of sand bed grain sizes. The assumption that each particle size saltates independently of other particles, combined with the fact that both the fluid threshold for initiation and the impact threshold for cessation of saltation increase with particle size (e.g., Bagnold, 1937; Chepil, 1945b; Kok et al., 2012; Martin and Kok, 2018), implies size-selective susceptibility to saltation for different particle sizes on a mixed sediment bed (Figs. 1a and b). In contrast to this expectation of size-selective thresholds for the independent saltation approach, our observations indicate that a single threshold governs the occurrence of saltation for most particle size classes. This further indicates that different particle sizes are strongly interacting, which contradicts the fundamental assumption underpinning independent saltation models. It remains possible, however, that the coarse tail of the particle-size distribution may still be subject to the size-selective susceptibility of the independent saltation model. These coarse, low-energy saltators experience hops or rolling motions that do not rise far above the bed surface, and thus are not captured by our measurements. Such coarse tail particles may be transported instead primarily through creep, reptation, and weak saltation not captured by our measurements.

In contrast to independent saltation models, representative saltation models correctly predict the presence of a common threshold stress $\tau_{th}$ for the susceptibility of most particle size classes to saltation (Fig. 3), suggesting that representative saltation models are a useful tool for determining when saltation will occur. Indeed, in a separate study (Martin and Kok, 2018), we used $d_{50,bed}$ as the representative soil particle size governing the saltation threshold at our field sites, and we found bulk fluid and impact threshold stresses similar to those obtained for the same particle diameters in single particle wind tunnel and numerical studies (e.g., Bagnold, 1937; Chepil, 1945b; Kok et al., 2012). Wind tunnel experiments (Chepil, 1945b; Nickling, 1988) further show that the bulk saltation threshold for both uniform and bimodal sediment beds is determined by the mean bed grain size $\bar{d}_{bed}$; similarly, field observations indicate that bulk threshold is controlled by modal grain size $d_{modal,bed}$ (Gillette et al., 1980). Other experiments with bimodal bed sediments contrastingly report bulk saltation thresholds that substantially exceed threshold expectations for the median bed grain size $d_{50,bed}$ (McKenna Neuman and Bédard, 2017), which is consistent with the finding that bulk threshold increases with the areal coverage of coarse (i.e., > 1mm) particles on a soil surface (Gillette et al., 1980). Further work is thus needed to better understand how the surface PSD, and other factors such as the configuration of coarse particles (e.g., Greeley et al., 1974; Iversen et al., 1976; Gillette and Stockton, 1989) and availability of loose sediment (e.g., Sharp, 1964; Gillette and Chen, 2001), determine both bulk and size-selective saltation thresholds.

### 4.4.2. Applications of equal susceptibility and size-selective mobility for modeling aeolian processes on Earth and planetary surfaces

Our results suggest that a representative particle size, possibly $d_{50,bed}$, could be used to determine the saltation threshold for most particle sizes on a natural mixed sediment bed, and that a representative particle size could also potentially be used to estimate the bulk coefficient (but possibly not size-selective coefficients) for the saltation flux law (Eq. 1). This representative saltation approach roughly parallels the adoption of a reference grain size for studies of water-driven bedload transport (e.g., Wiberg and Smith, 1989; Parker, 1990), for which equal susceptibility has also been observed (e.g., Andrews, 1983; Wiberg and Smith, 1987). However, such fluvial bedload differs from aeolian saltation in that fluvial particle entrainment is dominantly governed by fluid lifting (e.g., Wiberg and Smith, 1987) rather than by particle impacts, as for aeolian saltation (e.g., Pähtz and Durán, 2017).

On Earth (e.g., Gillette et al., 1980), Mars (e.g., Sullivan et al., 2008; Lapotre et al., 2016), and other planetary surfaces (e.g., Claudin and Andreotti, 2006; Lorenz and Zimbelman, 2014) containing widely mixed soil PSDs, the ability to choose a representative grain size for saltation could simplify the modeling of aeolian geomorphic processes, such as the migration of dunes and ripples (e.g., Bridges et al., 2012). However, equal susceptibility in saltation is not universal; in our case, it is limited only to particles below a certain size, $d_i/d_{50,bed} \approx 1.5$. For highly heterogeneous or bimodal sediment beds, the coarse tail of the soil PSD may be mobilized primarily in small reptating or "creeping" motions driven by the impacts of finer particles (e.g., Bagnold, 1937; Ungar and Haff, 1987; Namikas, 2003; Yizhaq et al., 2012; Cheng et al., 2015), with important implications for segregation of different grain sizes on the sediment bed (e.g., Bagnold, 1941; Anderson and Bunas, 1993; Jerolmack et al., 2006; Isenberg et al., 2011). Related to this, sediment beds containing a very wide distribution of particle sizes may form coarse armors that substantially inhibit the occurrence of saltation (Gao et al., 2016; McKenna Neuman and Bédard, 2017). Size-selective mobility can also cause spatial sorting of different particle size classes, especially over irregular terrain (e.g., Arens et al., 2002). Such armoring and sorting effects may explain why particle sizes in aeolian deposits appear to vary with mean wind speed (Lindhorst and Betzler, 2016), despite the invariance in the distribution of transported sediment sizes suggested by equal susceptibility.

Our results can also help inform improved parameterizations of mineral dust emission used in atmospheric circulation models. Because dust emission is governed by the rate and speed of saltator impacts with the soil bed (e.g., Shao, 2008; Kok et al., 2014), it is dependent on the PSD of airborne saltators (e.g., Shao et al., 1993; Alfaro and Gomes, 2001). However, in the absence of size-selective sand transport measurements, most dust emission models have adopted the independent saltation approach, assuming that each particle size class in the sediment bed saltates as if it were the only particle size (Marticorena and Bergametti, 1995; Shao et al., 1996; Alfaro and Gomes, 2001; Zender et al., 2003). This has led to important inaccuracies in dust emission models. In particular, such dust emission models predict that the size distribution of saltating sand grains depends on the shear stress, which we find to be mostly incorrect (Fig. 3). This stress-dependence of modeled saltator PSDs contributes to model-predicted fining of emitted dust with increasing wind strength, which is indeed inconsistent with a compilation of field measurements (Kok, 2011) and recent wind tunnel results (Parajuli et al., 2016). In fact, these recent measurements show that the airborne dust size distribution is independent of wind shear stress, in analogy to our finding here that the mobility of sand-sized particles is not substantially dependent on wind stress (Fig. 3). By rejecting the independent saltation approach, these dust emission models could thus be improved by instead assuming that the distribution of saltation impact speeds is governed by a single representative grain size. However, this approach also has limitations, because the ejection of dust aerosols from the soil bed depends on the distribution of energies of particle impacts (e.g., Alfaro and Gomes, 2001; Shao, 2008; Kok et al., 2014), which in turn depends directly on the full PSD of impacting saltators. To inform such improved dust emission models, further work is needed to provide trustworthy estimates of size-selective mobility (i.e., Fig. 4) and to parameterize the relationship between the distribution of saltation impact speeds and the PSD of the soil surface.

## 5. Conclusions
We have used extensive field measurements to show that aeolian saltation over a natural soil bed containing a mixture of sediment sizes displays equal susceptibility to saltation for particle sizes of at least ~1.5 times the median bed particle diameter. This invariance to particle size of the threshold wind stress needed to entrain particles appears to result from the fact that saltation impact speeds, which drive particle entrainment, are mostly insensitive to the wind shear stress.

We also explored the possibility that the mobility of different particle sizes is size-selective, but our interpretations of this behavior were limited by the lack of near-surface measurements of airborne particle size distributions.

These findings refute treatments of saltation and associated dust emission that assume independent saltation dynamics for each particle size class, whereas they provide qualified support for the adoption of a single representative bed particle size for determining the saltation threshold and the bulk sand flux. Adopting such a representative grain size for modeling aeolian saltation on natural mixed sediment beds could help to both simplify and improve predictions of geomorphic evolution and dust emission on Earth and planetary surfaces.

# Appendix A. Calculation of combined airborne particle size distributions (PSDs)

To calculate a vertically-integrated airborne PSD $\frac{dV_{air}}{d\ln(d)}$ for each ~1 hr BSNE saltation trap measurement interval, we compute a weighted sum of individual PSDs:

$$\frac{dV_{air}}{d\ln(d)} = \sum_j \frac{dV_j}{d\ln(d)} w_j, \quad (A1)$$

where $\frac{dV_j}{d\ln(d)}$ is the PSD for BSNE $j$, and $w_j$ is a weight applied to this sample, such that $\sum_j w_j = 1$ for the combination of all BSNEs in the profile. At BSNE heights for which we obtained PSDs for combined samples, we set $\frac{dV_j}{d\ln(d)}$ as the PSD for the combined sample containing the BSNE measurement interval for the profile of interest. In certain cases (e.g., days with low saltation flux), there are insufficient airborne PSDs to adequately calculate $\frac{dV_{air}}{d\ln(d)}$. To ensure that $\frac{dV_{air}}{d\ln(d)}$ adequately represents the airborne PSD over a wide range of heights, we limit analyses only to those time intervals containing at least 3 airborne PSDs, at least one PSD with $z \leq 2.5 z_q$, and at least one PSD with $z \geq 4.5 z_q$ (Supporting Information Fig. S3).

We compute $w_j$ values by estimating the contribution of each BSNE height $z_j$ to the exponential profile that characterizes the variation of the horizontal flux with height (Martin and Kok, 2017). Consequently, each measured PSD is weighted by its contribution to the vertically-integrated saltation flux. Adapting the weighting approach of Martin et al. (2018; see Sec. 4.4.2 for details), we compute each weight as the relative incremental contribution of the range of heights covered by a single BSNE trap to the calculation of the total vertically-integrated saltation flux:

$$w_j = \frac{\exp(1)}{z_q}\left[\exp\left(-\frac{z_{j,bot}}{z_q}\right) - \exp\left(-\frac{z_{j,top}}{z_q}\right)\right], \quad (A2)$$

where $z_q$ is a best-fit $e$-folding height determined by fitting an exponential profile to the bulk (non-size-selective) saltation flux profile, i.e. $q(z) = q_0 \exp(-z/z_q)$ (Martin et al., 2018). The heights $z_{j,bot}$ and $z_{j,top}$ indicate the spatial coverage for each BSNE within an exponential profile. We therefore calculate each $z_{j,bot}$ and $z_{j,top}$ as a geometric mean with respect to adjacent BSNE heights: $z_{j,bot} = \sqrt{z_j z_{j-1}}$ and $z_{j,top} = \sqrt{z_j z_{j+1}}$. Here, $z_{j-1}$ is the height of the BSNE trap below $z_j$, and $z_{j+1}$ is the height of the BSNE trap above $z_j$. If $z_j$ is the highest BSNE in the profile, then we set $z_{j,top} = \infty$. If $z_j$ is the lowest BSNE in the profile, then we set $z_{j,bot} = z_q$. We set this lower limit for the weighting profile as $z_q$ (not $z = 0$); because nearly all BSNE traps were situated above $z_q$ (see Supporting Information Fig. S3). To account for this lower limit for computing $w_j$, we include the $\exp(1)$ term in Eq. A2 to ensure that the sum of the $w_j$ equals 1. Therefore, as we consider in Sec. 4.1 of the main text, our calculated vertically-integrated airborne PSDs $\frac{dV_{air}}{d\ln(d)}$ represent only the upper part, i.e., $z > z_q$, of the saltation profile.

**ACKNOWLEDGEMENTS.** U.S. National Science Foundation (NSF) Postdoctoral Fellowship EAR-1249918 to R.L.M. and NSF grant AGS-1358621 to J.F.K. supported this research. Research was also sponsored by the Army Research Laboratory and was accomplished under Grant Number W911NF-15-1-0417. The views and conclusions contained in this document are those of the authors and should not be interpreted as representing the official policies, either expressed or implied, of the Army Research Laboratory or the U.S. Government. The U.S. Government is authorized to reproduce and distribute reprints for Government purposes notwithstanding any copyright notation herein. Oceano Dunes State Vehicular Recreation Area, Rancho Guadalupe Dunes Preserve, and Jericoacoara National Park provided essential site access and support. Jericoacoara fieldwork is registered with the Brazilian Ministry of the Environment (#46254-1 to J. Ellis). We thank Marcelo Chamecki for advice on treatment of wind data, Chris Hugenholtz and Tom Barchyn for equipment help, Doug Jerolmack for lab access for grain-size analysis, and Jean Ellis, Paulo Sousa, Peter Li, Francis Turney, Arkayan Samaddar, and Livia Freire for field assistance. Data included in the analysis for this paper can be found on the Zenodo data repository at http://doi.org/10.5281/zenodo.2631823.

Supporting Information for

# Size-independent susceptibility to transport in aeolian saltation

Raleigh L. Martin[1] and Jasper F. Kok[1]

[1]Department of Atmospheric and Oceanic Sciences, University of California, Los Angeles, CA 90095

## Contents of this file

Figures S1-S5

## Introduction

In this supporting document, we provide additional figures to give further justification for methods and results described in the main paper.

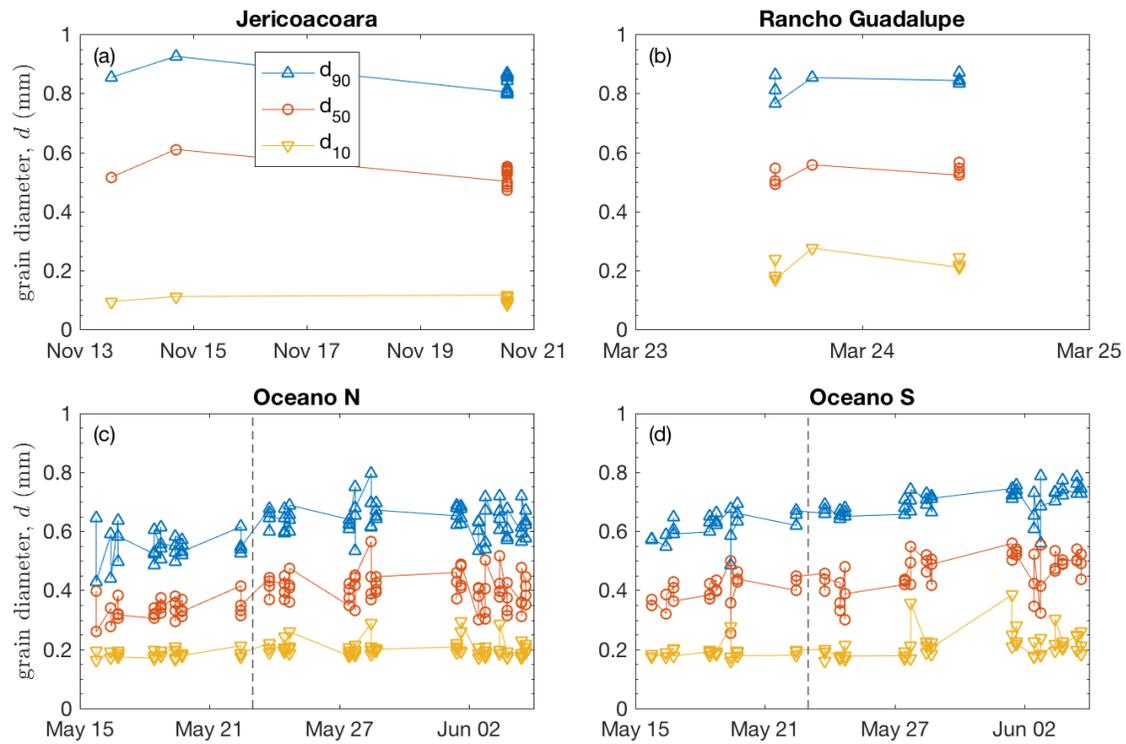

**Figure S1.** Variation in 10th percentile ($d_{10}$), 50th percentile ($d_{50}$), and 90th percentile ($d_{90}$) grain diameter of individual bed surface sediment samples at each field site. Data at Oceano are separated into the northern cluster ("N") and southern cluster ("S") of BSNE saltation traps where bed surface samples were collected. Dashed line at Oceano indicates separation between samples for first half of deployment (May 15 – May 22, 2015) and second half of deployment (May 23 – June 4, 2015), which we analyzed separately.

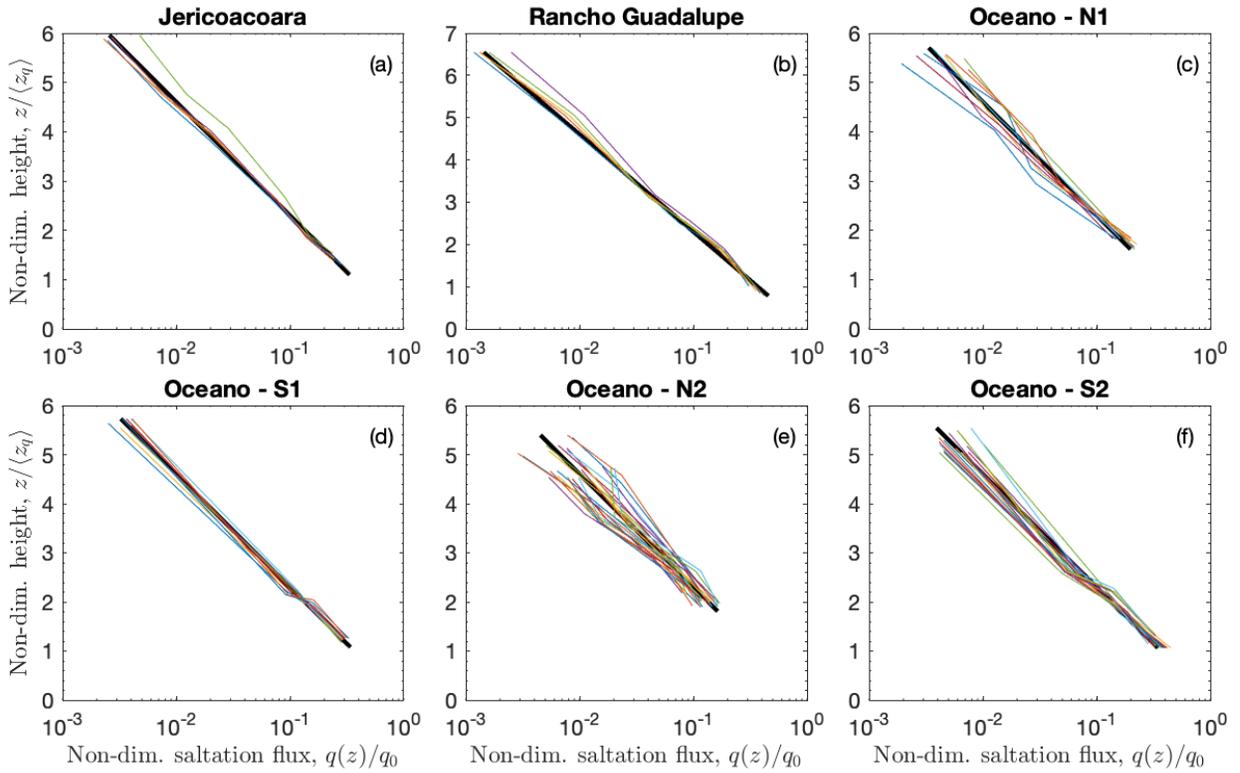

**Figure S2.** Individual saltation flux profiles $q(z)$ compared to idealized exponential profiles. For each site, height-specific fluxes $q$ are non-dimensionalized by the profile fitting parameter $q_0$, and heights $z$ are non-dimensionalized by the mean $e$-folding height $\langle z_q \rangle$. Solid black lines show idealized exponential profiles, i.e.: $q(z) = q_0 \exp(-z/z_q)$. Methods for obtaining $q_0$ and $z_q$ are described in Martin et al. (2018).

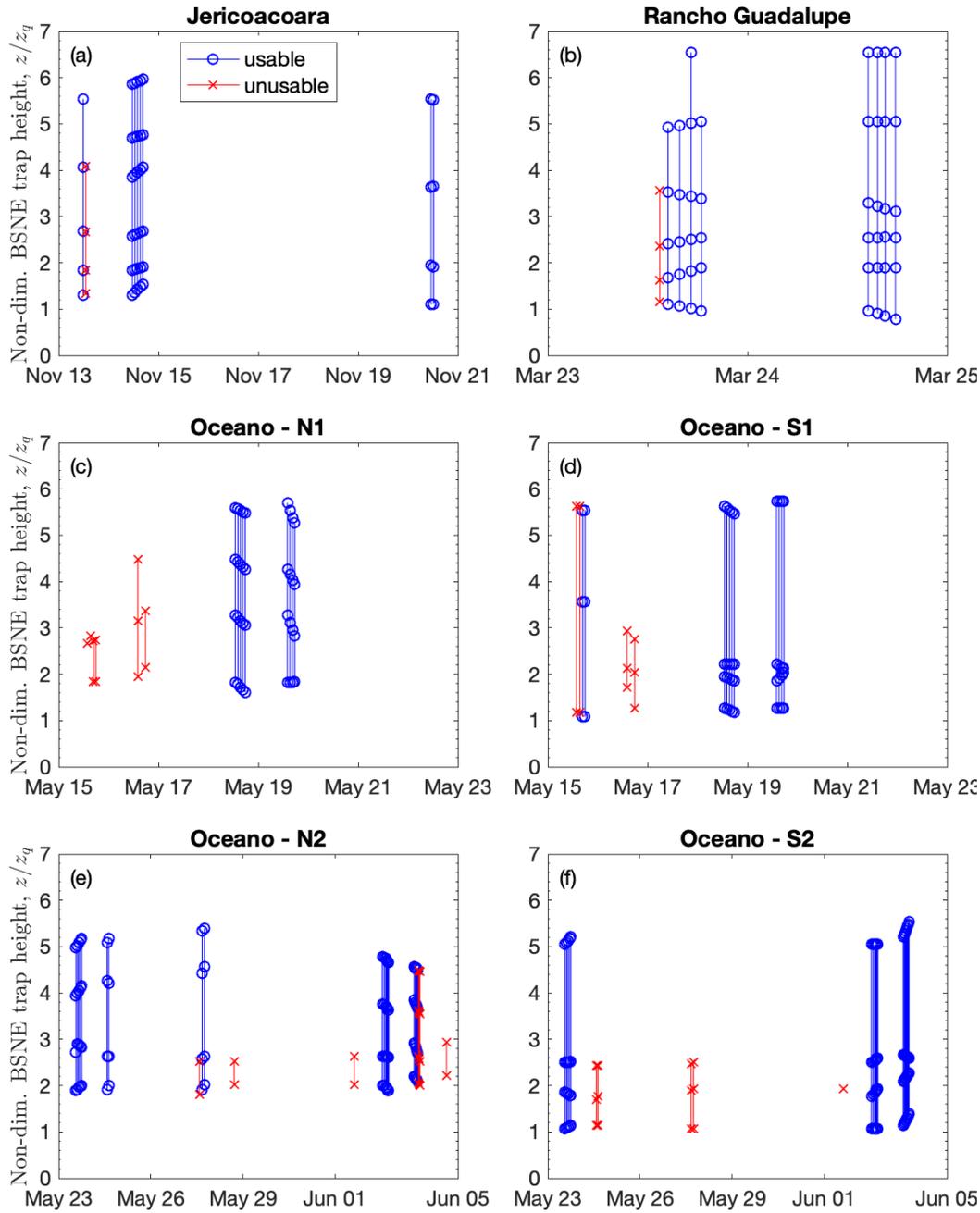

**Figure S3.** Profiles of BSNE saltation trap heights $z$ associated with airborne PSD measurements, non-dimensionalized by campaign-averaged $e$-folding saltation layer heights $z_q$ for each site (Martin and Kok, 2017). Certain days (e.g., May 22, 2015, at Oceano) lack airborne PSDs despite the presence of surface PSDs. On these days, saltation fluxes were too small to provide sufficient airborne samples for any PSD measurements.

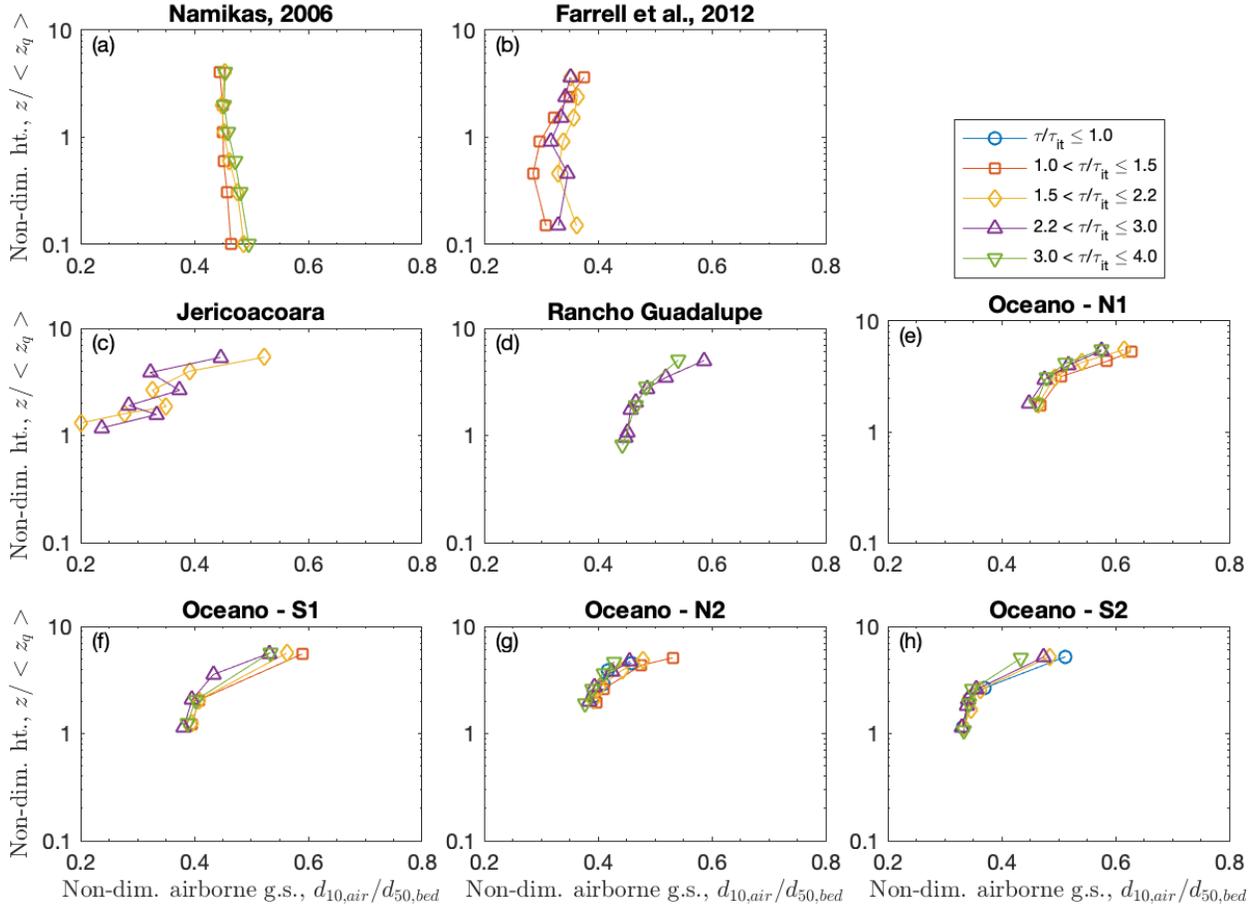

**Figure S4.** 10[th] percentile airborne grain diameter $d_{10,air}$ versus height above the bed surface $z$, for measurements (a) by Namikas (2006), (b) by Farrell et al. (2012), and (c-h) at our field sites. $d_{10,air}$ is non-dimensionalized by median bed grain diameter $d_{50,bed}$, and $z$ is non-dimensionalized by mean $e$-folding saltation layer height $\langle z_q \rangle$. Measurements are further subdivided by shear stress $\tau$ non-dimensionalized by impact threshold stress $\tau_{it}$. Values for $d_{50,bed}$, $z_q$, and $\tau_{it}$ for Namikas and Farrell et al. were obtained as described in the caption to Fig. 6 in the main text. Farrell et al. further did not report full airborne PSDs or $d_{10,air}$, thus we calculate $d_{10,air}$ based on the mean and standard deviation airborne grain diameters reported in their study and assuming lognormal PSDs.

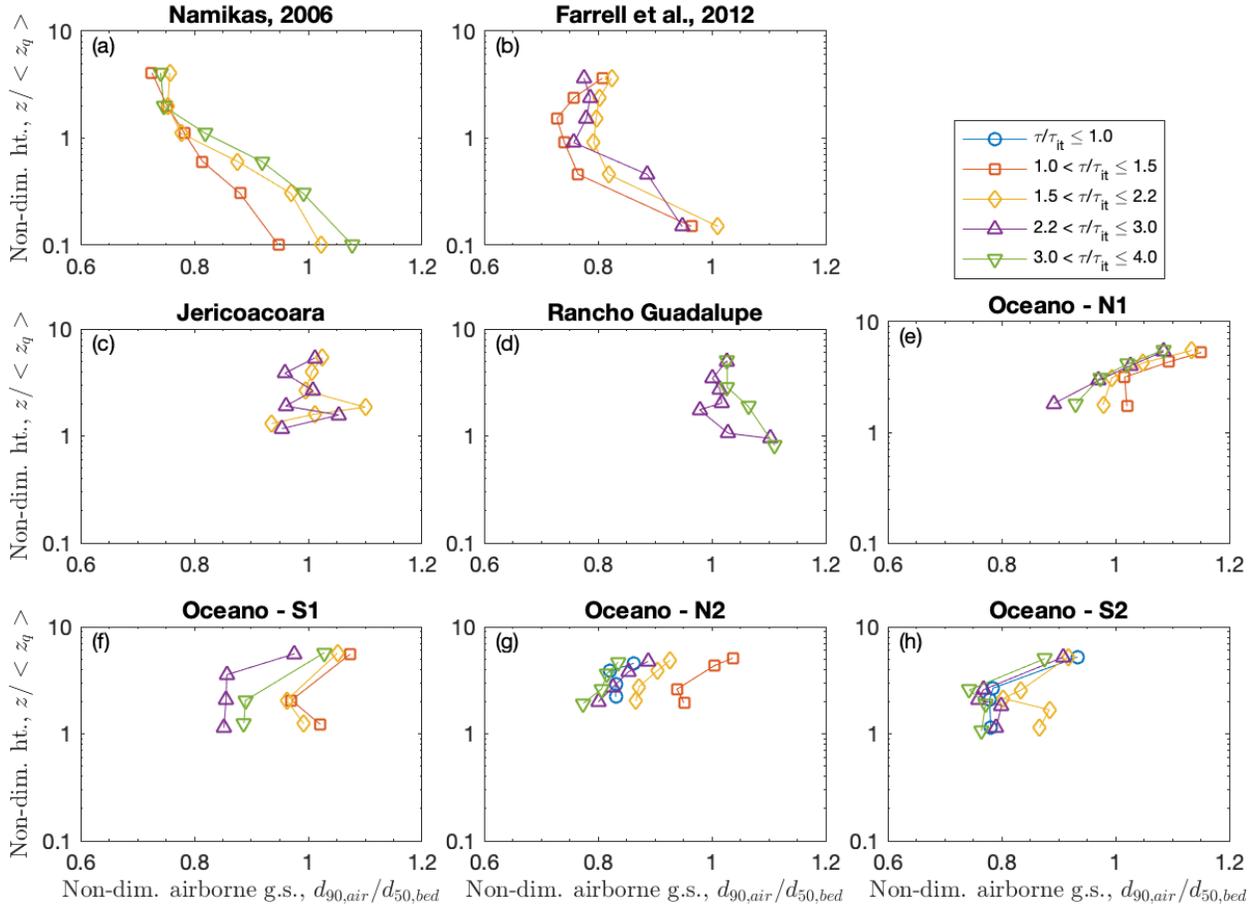

**Figure S5.** 90[th] percentile airborne grain diameter $d_{90,air}$ versus height above the bed surface $z$, for measurements (a) by Namikas (2006), (b) by Farrell et al. (2012), and (c-h) at our field sites. $d_{90,air}$ is non-dimensionalized by median bed grain diameter $d_{50,bed}$, and $z$ is non-dimensionalized by mean $e$-folding saltation layer height $\langle z_q \rangle$. Measurements are further subdivided by shear stress $\tau$ non-dimensionalized by impact threshold stress $\tau_{it}$. Values for $d_{50,bed}$, $z_q$, and $\tau_{it}$ for Namikas and Farrell et al. were obtained as described in the caption to Fig. 6 in the main text. Farrell et al. further did not report full airborne PSDs or $d_{10,air}$, thus we calculate $d_{10,air}$ based on the mean and standard deviation airborne grain diameters reported in their study and assuming lognormal PSDs.